\documentclass{camera}
\newcommand{\smartpap}{p\hskip-7pt\hbox{$^{^{(\!-\!)}}$}}
\usepackage{graphicx}
\begin{document}
\title{ELECTROWEAK RADIATIVE CORRECTIONS AND THE $W$ BOSON MASS AT HADRON
COLLIDERS}
\author{C.M. Carloni Calame$^{1,2}$, G. Montagna$^{2,1}$,
O. Nicrosini$^{1,2}$ \and M. Treccani$^2$}
\organization{$^1$ INFN, Sezione di Pavia \\ 
$^2$ Dipartimento di Fisica Nucleare e Teorica, 
Universit\`a di Pavia \\ via A. Bassi 6, I - 27100, Pavia, Italy}
\maketitle
\noindent
At present and future hadron colliders, the precision physics program started
in the past will be continued. In particular, a precise determination of the
$W$ boson mass will be carried out. This requires the calculation of the 
radiative corrections and their implementation in Monte Carlo event
generators for data analysis. In this talk, 
the status of the calculation of the order $\alpha$ electroweak
radiative corrections is reviewed and a study of the impact of higher order
QED corrections on the $W$ boson mass is presented.
\section{Introduction}
\label{intro}
In addition to the program of discovery physics, experiments at the high-energy
hadron colliders Tevatron RunII and the LHC are expected to continue the
program of precision physics successfully carried out at LEP, SLC and the
Tevatron itself. In particular, for precision tests of the Standard Model a
very precise determination of the $W$ boson mass $M_W$ is important because
this, together with an improved measurement of the top-quark mass, will put
more severe indirect bounds on the mass of the Higgs boson.

The precision expected at the Tevatron is about 30 MeV 
per experiment per channel at Run IIa and 16 MeV at Run IIb, 
the latter being the same precision aimed at at the LHC~\cite{exp}.
Therefore, precise calculations and
event generators for the Drell-Yan process $p\smartpap \to W \to l 
\nu_l$ 
are strongly required.

%
From the experimental point of view,
the $W$ mass is extracted from the kinematics 
of the $W$ boson decay $W \to l \nu_l$,
giving rise to a Jacobian peak in the distribution of the lepton transverse
momentum $p_T(l)$. However, the preferred quantity to determine $M_W$ 
is the transverse mass spectrum $M_T$~\cite{exp}, 
which is less
sensitive than the $p_T(l)$ distribution to the $W$ transverse motion.

Having in mind the precision anticipated at
the Tevatron and the LHC for $M_W$, accurate theoretical predictions including
QCD and electroweak (EW) radiative corrections (RC) are necessary to precisely
extract $M_W$ from the data. In the following, we concentrate on EW RC.

\section{Order $\alpha$ electroweak radiative corrections}

In the past years, the calculation of the full set of the $\cal{O}(\alpha)$ EW
RC to the single-$W$ production in hadronic collisions was carried out. The
first calculations were performed in the resonant $W$ approximation (pole
approximation) in Refs. \cite{wh,bkw} and then the complete
$\cal{O}(\alpha)$ corrections were calculated in Ref. \cite{dk}. It is
worth noticing that the non-resonant contributions are important in the tail
of the $M_T$ distribution, far from the $M_W$ peak, because large Sudakov-like
logarithms arise. Since  the $W$ width can be measured in the $M_T$ tail, the
full calculation is then mandatory.

In the present experimental analyses, the corrections of Ref. \cite{bkw} are
included. It comes out that EW corrections
shift the $W$ mass 
(in the measurements at 
the Tevatron Run Ib) 
by an amount of 
$-65 \pm 20$ MeV and $-168 \pm 10$ MeV for electron and muon channels,
respectively \cite{exp}. It is known that
these shifts are mainly due to final-state photonic corrections because of the
presence of large collinear logarithms. 
In the presence of realistic selection criteria, 
the correction due to final-state photon radiation is of several per cent 
on the $M_T$ spectrum in the peak region $M_T \approx M_W$.
This poses the question of the impact of higher-order 
QED corrections due to the multiple emission of (real and virtual) photons.
These higher-order contributions are not presently included in data analysis
at the Tevatron but they are estimated to introduce a systematic uncertainty of
20 MeV in the $W \to e \nu_e$ decay channel and 10 MeV in the $W \to \mu
\nu_\mu$ decay \cite{exp}.
This source of systematic uncertainty is not negligible in view of the foreseen
experimental precisions and it can be reduced by means of improved theoretical
calculations.
Recent work in this direction includes 
the calculation of multi-photon corrections to leptonic $W$ decays in the
framework of the 
YFS approach 
implemented in the Monte Carlo (MC) generator $\tt WINHAC$ \cite{winhac}.

\section{Higher-order QED corrections}
In the approach here presented the real plus virtual corrections due to
multi-photon radiation are computed in the leading-log approximation using
the QED structure-function approach. The
corrections are calculated by solving the QED DGLAP equation by means of the
QED Parton Shower (PS) algorithm
developed in Ref.~\cite{ps}. Only radiation from the final-state leptons 
is presently included in our approach, 
because it is known that quark-mass singularities, originating from
initial-state photon radiation, can be reabsorbed into a redefinition
of the Parton Distribution Functions \cite{sp}. 
After this mass-factorization procedure, initial-state-radiation
has only a small and uniform impact on the $M_T$ spectrum, while final-state
radiation
significantly distorts the shape of the $M_T$ distribution, affecting
in turn the $M_W$ extraction. 

The formulation is implemented into a MC generator, 
$\tt HORACE$ \cite{cmnt}, which incorporates lepton identification criteria
and detector resolution effects, in order to perform simulations
for the hadronic process $p \smartpap \to W \to l \nu$ ($l=e,\mu$) 
as realistic as possible. $\tt HORACE$ calculates QED corrections to all
orders and at order $\alpha$, 
to disentangle the effect of higher-order contributions and to compare with
the available $O(\alpha)$ programs. A first comparison was performed
with $\tt WGRAD$ \cite{bkw}, showing good agreement. 
A more detailed analysis, comparing $\tt HORACE$ and $\tt WINHAC$, is in
progress \cite{jpcmn}.

To evaluate the shift on the fitted $W$ mass induced by higher-order QED
corrections we used $\tt HORACE$ and performed $\chi^2$ fits to MC pseudo-data 
for the $M_T$ spectrum, simulating acceptance cuts, lepton identification
criteria and (simplified) detector resolution effects. The center of mass
energy considered
in our study is $\sqrt{s} = 2$ TeV, corresponding to the Tevatron, but we 
checked that the conclusions of our analysis do not significantly change at
the LHC energy. 
The steps of the fitting procedure are described in detail in Ref.~\cite{cmnt}.
The results of our analysis show that the
mass shift due to higher-order effects is about $10$~MeV for the
$W\to\mu\nu$ channel and a few MeV for the $W \to e \nu$
channel, as a consequence of the different identification requirements for
electrons and muons. Therefore, in view of the
expected precision of 15-30 MeV for $M_W$ at the Tevatron Run II and at the
LHC, it will be important to take multi-photon effects into account when
extracting $M_W$ from the data.

\section{Conclusions}
In view of improved precision measurements of the $W$ mass at present and
future hadron colliders, accurate theoretical predictions including
EW radiative corrections are strongly required for data
analysis.

In recent years, the calculation of the full set of the $\cal{O}(\alpha)$
EW RC was carried out \cite{bkw,dk}. They are
currently included in the experimental analyses and induce a shift on the
extracted $W$ mass of the order of 100 MeV. The shift is mainly due to the
photonic contributions.

A theoretical systematic uncertainty which can limit the precision of the
future
measurement is the effect of multi-photon emission. We studied the problem
within a QED PS approach, by means of the event generator
$\tt HORACE$.
%
We found that the
shift due to these corrections is about 10~MeV in the $W \to \mu \nu$ channel
and a few MeV in the $W \to e \nu$ channel. Therefore, they are important in
view of the aimed precision of the order of 15-30 MeV.

\section{Acknowledgments}
The authors would like to thank the IFAE 2003 organizers and the conveners of
the electroweak session for the organization of the wide and interesting
scientific programme.
%
%

%
\end{document}